\documentclass[10pt, twocolumn, pre, aps, superscriptaddress, showpacs]{revtex4-1}
\usepackage{amsmath, graphicx, subfigure, tikz}
\begin{document}
    \title{Frustrated Potts: Multiplicity Eliminates Chaos via Reentrance}
    \author{Alpar T\"urko\u{g}lu}
    \affiliation{Department of Electrical and Electronics Engineering, Bo\u{g}azi\c{c}i University, Bebek, Istanbul 34342, Turkey}
    \author{A. Nihat Berker}
    \affiliation{Faculty of Engineering and Natural Sciences, Kadir Has University, Cibali, Istanbul 34083, Turkey}
    \affiliation{Department of Physics, Massachusetts Institute of Technology, Cambridge, Massachusetts 02139, USA}

    \begin{abstract}
The frustrated $q$-state Potts model is solved exactly on a
hierarchical lattice, yielding chaos under rescaling, namely the
signature of a spin-glass phase, as previously seen for the Ising
$(q=2)$ model. However, the ground-state entropy introduced by the
$(q>2)$-state antiferromagnetic Potts bond induces an escape from
chaos as multiplicity $q$ increases.  The frustration versus
multiplicity phase diagram has a reentrant (as a function of
frustration) chaotic phase.
    \end{abstract}
    \maketitle

Frustration \cite{Toulouse}, meaning loops of equal-strength
interactions that cannot all be simultaneously satisfied, diminishes
ordered phases in the phase diagram and may drastically change the
nature of certain ordered phases.\cite{Gulpinar}  For example, the
so-called Mattis phase \cite{Mattis}, where spins are ordered in random directions
but the system has no competing interactions between the spins and thus all interactions can be simultaneously satisfied, becomes a
spin-glass phase with the introduction of the smallest amount of competing frustrated interactions  \cite{Ilker2}, with residual entropy, unsaturated order at zero temperature, and the chaotic rescaling of the interactions,
as measured by a positive Lyapunov exponent. The latter, chaos under
scale change, is the signature of the spin-glass phase
\cite{McKayChaos,McKayChaos2,BerkerMcKay,Hartford,ZZhu,Katzgraber3,Fernandez,Fernandez2,Eldan},
as gauged quantitatively by the Lyapunov exponent. Chaotic
interactions under scale change dictate chaotic correlation
functions as a function of distance.\cite{Aral}

\begin{figure}[ht!]
\centering
\includegraphics[scale=0.4]{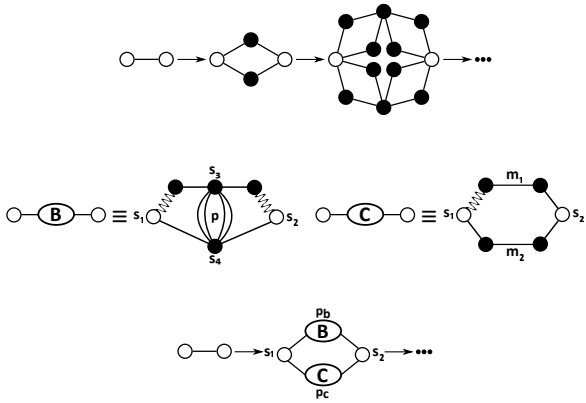}
\caption{ Top row: Construction of a hierarchical lattice, from
Ref.\cite{BerkerOstlund}. The renormalization-group solution of a
hierarchical lattice proceeds via renormalization group in the
opposite direction of its construction. Middle row: The two units
used in the construction of the frustrated hierarchical
lattice.\cite{McKayChaos} On the left is the frustrated
unit and on the right the repressed unit.  In the former, $p$ parallel bonds connect the internal spins $s_3$ and $s_4$ ($p=4$ is shown here). The wiggly bonds are
infinitely strong antiferromagnetic couplings. Bottom row: The
assemblage of the units for the construction of the frustrated
hierarchical lattice.}
\end{figure}

In fact, chaos under rescaling was seen in frustrated systems, with
no randomness, with the exact solution of hierarchical
lattices \cite{McKayChaos,McKayChaos2,BerkerMcKay,Hartford}.
When frustration is increased, a sequence of period doublings leads to the regime of spin-glass chaos \cite{McKayChaos}.  The regime of period doublings that precedes spin-glass chaos also converts, under randomness, to chaotic bands \cite{McKayChaos2}. Spin-glass chaos and its positive
Lyapunov exponent was also calculated in the renormalization-group
solution of cubic systems where frustration is introduced by quenched
randomness.\cite{Ilker2}

\begin{figure}[ht!]
\centering
\includegraphics[scale=0.07]{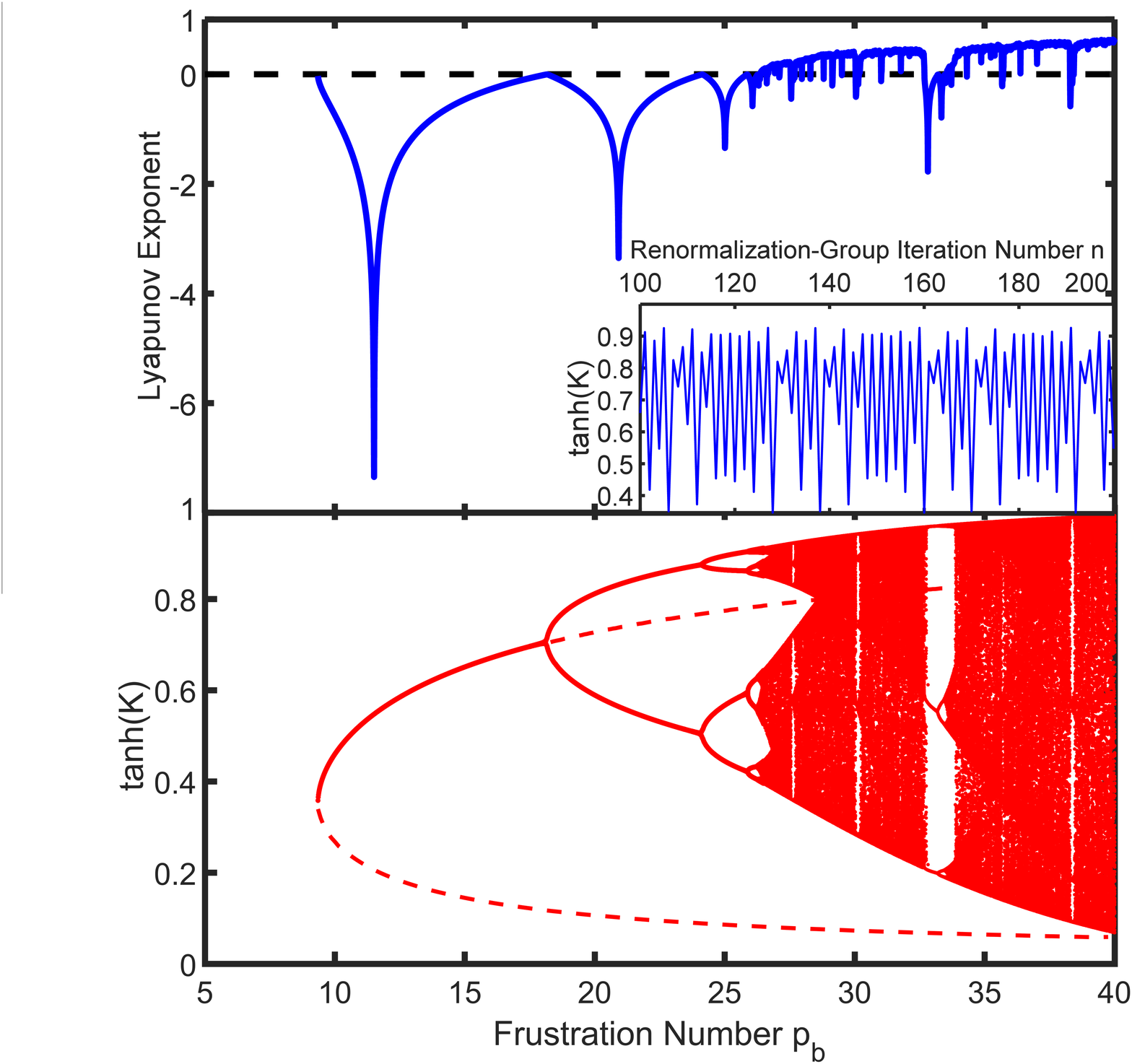}
\caption{Lower panel: The onset of chaos, by period doubling, under
increased frustration for the $q=3$-state Potts model. For each
$p_b$, the renormalization-group flows are in the vertical
direction.  The full lines represent attractive fixed points, limit
cycles, and chaotic bands.  The dashed lines represent the unstable
fixed points, only some of which are shown. The $tanh(K) = 0$ points
are the stable fixed points which are the sinks of the disordered
phase. The upper panel shows the calculated Lyapunov exponents. The
upper inset shows the chaotic renormalization-group trajectory for
$p_b = 28.$}
\end{figure}

Another important cause of ground-state degeneracy is in the
ground-state participation of a multiplicity of spin states, as
seen in antiferromagnetic $(q>2)$-state Potts
models \cite{BerkerKadanoff1,BerkerKadanoff2,Saleur}.  In the latter systems, antiferromagnetic Potts interactions can all be satisfied, while local spins can take many different values, leading to a distinctive phase \cite{BerkerKadanoff1,BerkerKadanoff2,Saleur}. In the current
work, we have studied the combination of both effects, chaos from
frustration and degeneracy from multiplicity of states. We have
exactly solved the $q$-state Potts models on the frustrated
hierarchical model as in Ref.\cite{McKayChaos}. As seen below and in Fig. 1, hierarchical models are models obtained by self-imbedding a graph infinitely many times and are exactly solvable by renormalization-group theory \cite{BerkerOstlund, Kaufman1, Kaufman2}.  We find that the
system escapes chaos through multiplicity and that chaos shows
reentrant behavior.  

\begin{figure}[ht!]
\centering
\includegraphics[scale=0.07]{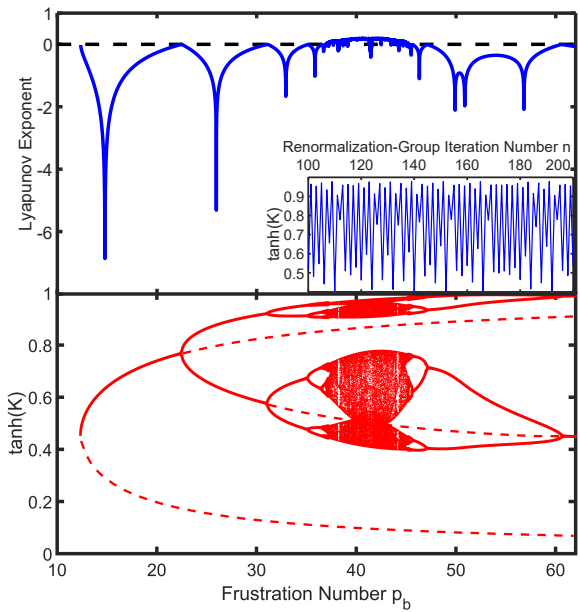}
\caption{The $q=6$-state Potts model, under increased frustration,
after the onset of chaos, leaves chaos through a set of reverse
period doublings. The upper inset shows the chaotic
renormalization-group trajectory for $p_b = 42.$}
\end{figure}

\begin{figure}[ht!]
\centering
\includegraphics[scale=0.07]{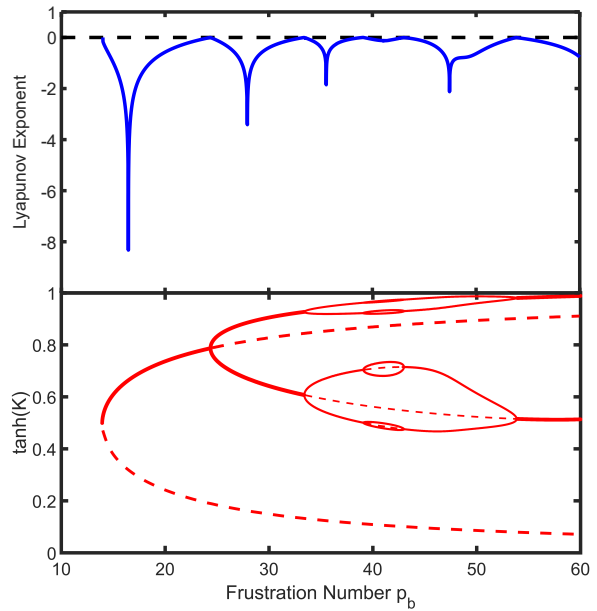}
\caption{The $q=8$-state Potts model, under increased frustration,
through a set of period doublings followed by reverse period
doublings, bypasses chaos.}
\end{figure}

\begin{figure}[ht!]
\centering
\includegraphics[scale=0.07]{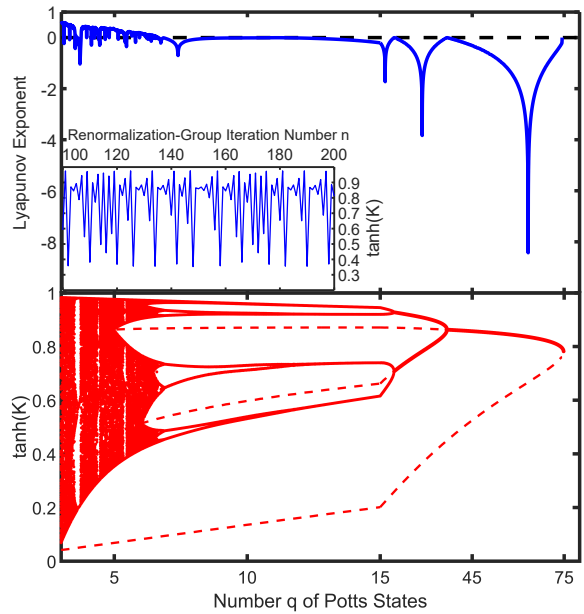}
\caption{As the number $q$ of Potts states increases, the system
leaves chaos through reverse period doublings.  Note the horizontal
scale change in the figure at $q=15$.  For this calculation,
$p_b=40$. The upper inset shows the chaotic renormalization-group
trajectory for $q=5$.}
\end{figure}

The model is constructed, by combining two units, embodying the
different microscopic effects of competing interactions.  The
Hamiltonian of the system is
\begin{equation}
- \beta {\cal H} = K \sum_{\left<ij\right>} \, \delta(s_i, s_j),
\end{equation}
where $\beta=1/k_{B}T$, at site $i$ the Potts spin $s_{i}=1,2,,...,q$ can
be in $q$ different states, the delta function $\delta(s_i,
s_j)=1(0)$ for $s_i=s_j (s_i\neq s_j)$, and the sum is over all
interacting pairs of spins, represented by straight lines in Fig. 1.
In unit $C$ of Fig. 1, the correlations are repressed but not eliminated along
the path of the unit, as $m_2 > m_1$ and the competing
correlation on the longer unit is weaker. Unit $B$ is frustrated: Both interaction loops across the unit are frustrated. The above are representative of the two qualitatively different effects of competing interactions in $d$-dimensional hypercubic lattices \cite{McKayChaos}.  This competition occurs when, in two different paths connecting two points on the lattice, all interactions cannot be simultaneously satisfied.

The recursion relations of a hierarchical lattice are obtained by doing the partition function sum of the spins internal to the unit (in black in Fig. 1) while keeping the two external spins (in white in Fig. 1) fixed \cite{BerkerOstlund, Kaufman1, Kaufman2}.  Thus, the two strands in unit C are disconnected with respect to this summation.   If the two paths are of unequal length, the effective overall interaction along the longer path is weaker (as can be seen, e.g., from renormalization-group transformations on $d=1$ chains), so that the overall interaction along the shorter path dominates but is weakened (repressed) by the parallel presence of the competing longer path.  For this, it is important that one of the paths has an anti-aligning wiggly bond, to induce competition rather than cooperation. On the other hand, when the two paths are of equal length, a frustrated loop is generated and the effective (\textit{i.e.}, renormalized) interactions along the two paths cancel out (are frustrated). Across unit B in Fig. 1 are two adjacent frustrated loops, each with one wiggly (anti-aligning) interaction, so that this unit is frustrated.  Incidentally, this discussion shows that loops are essential to chaotic spin-glass behavior.  Although the discussion here was given in context of $d$-dimensional hypercubic lattices, it is also applicable to scale-free networks of physical and social nature \cite{networks0,networks}.

By combining in parallel
$p_b$ and $p_c$ units $B$ and $C$, a family of models is created. In
this work, we present results for $m_1=2,m_2=3,p=4, p_c=1$ and varying the
number $p_b$ of frustrated units.  However, we have also explored a wide range of model parameters $m_1,m_2, p, p_c$ and find that our chaos and reentrance results are generic.  

\begin{figure}[ht!]
\centering
\includegraphics[scale=0.27]{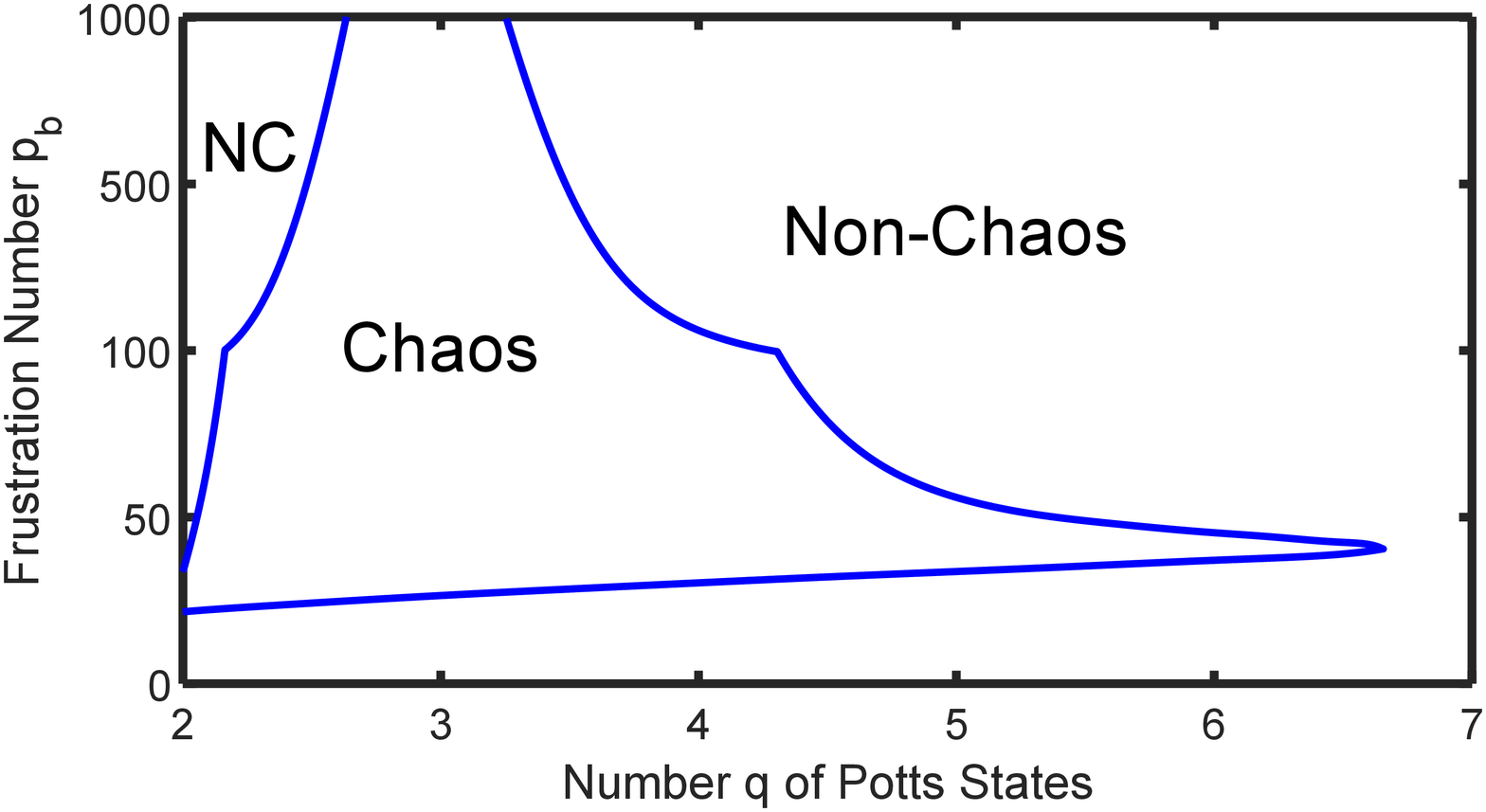}
\caption{The calculated frustration-multiplicity chaos phase diagram of the
system, shows the $(q,p_b)$ combinations for which chaos occurs.  NC
stands for non-chaos.  The vertical scale changes at $p_b=100$. Note
the reentrance, as a function of frustration, of non-chaos.  All the points of this phase diagram have been calculated.}
\end{figure}

Hierarchical models are solved exactly, by renormalization-group theory, proceeding in the reverse direction of the construction of the hierarchical model by decimating the interior spins of each level (black circles in Fig. 1), obtaining recursion relations $K' = K'(K)$
exactly.\cite{BerkerOstlund,Kaufman1,Kaufman2}. The onset of the chaotic bands, as the number $p_b$ of frustrated
units is increased, is shown in Fig. 2, for the number of Potts states $q=3$. The chaotic band is an interaction interval in which, under the renormalization-group transformations, the points are visited in a chaotic sequence.  It has been shown that chaotic interactions under renormalization-group dictate chaotic correlations between spins as a function of distance \cite{Aral}.  

A measure of the strength of chaos are the calculated Lyapunov exponents \cite{Collet,
Hilborn},
\begin{equation}
\lambda = \lim _{n\rightarrow\infty} \frac{1}{n} \sum_{k=0}^{n-1}
\ln \Big|\frac {dK_{k+1}}{dK_k}\Big|\,,
\end{equation}
where $K_k$ is the interaction at step $k$ of the
renormalization-group trajectory.  Thus, the Lyapunov exponent is calculated by logarithmically summing \cite{Collet,
Hilborn} the derivative $dK_{k+1}/d K_k$  of the interaction constant $K_{k+1}$ after the $(k+1)$th renormalization, with respect to the interaction constant  $K_k$ after the $(k)$th renormalization. The sum in Eq.(2) is to be taken
within the asymptotic trajectory, so that we throw out the first 100
renormalization-group iterations to eliminate the transient points
and subsequently use 600 iterations in the sum in Eq.(2), which
assures convergence in the chaotic bands. It is seen that the
Lyapunov exponent is non-positive outside chaos, barely touching
zero at each period doubling.  

A qualitative understanding of this chaos phenomenon is that strong interactions when frustrated generate weak interactions at the next length scale of the renormalization-group transformation.  Thus, renormalization-group flows inside the ordered phase cannot go to a strong-coupling sink \cite{BerkerWortis} as happens in non-frustrated systems.  This mechanism and phenomenon occur in hierarchical as well as non-hierarchical lattices, as is seen in in hypercubic lattices \cite{Ilker2} and as was eventually accepted (subsequently to \cite{McKayChaos}) for spin-glass phases in general \cite{McKayChaos,McKayChaos2,BerkerMcKay,Hartford,ZZhu,Katzgraber3,Fernandez,Fernandez2,Eldan}.

For $q=6$ in Fig. 3, as frustration is increased, the
system leaves chaos through a series of reverse period doublings
(foldings).  We thus have non-chaos reentrance \cite{Cladis} around
the chaotic phase. For $q=8$, shown in Fig. 4, doublings and
foldings succeed, but chaos has disappeared. We can therefore look
for reverse bifurcations as the number of Potts states $q$ is
increased for fixed frustration. This is seen in Fig. 5, where
it is indeed seen that chaos disappears as the antiferromagnetic
degeneracy of the Potts model is increased by increasing $q$.

The complete frustration versus multiplicity phase diagram has been
calculated and is given in Fig. 6, clearly showing reentrance. Note
the stability of the chaotic phase around $q=3$.

The disappearance of the spin-glass phase under increased frustration has been seen in $q=2$ Ising models that are microscopically rewired, in $d=2,3$, by a renormalization-group study similar to our current work \cite{Ilker2}.  The full double reentrance as a function of frustration and $q$ needs be studied by other methods, such as Monte Carlo sampling \cite{Malakis}.  The ousting of frustrated chaos by the multiplicity of local states could have a relevance to the mathematical Ising-type modeling of societal collective behavior \cite{ColBeh}.  Such studies typically assign Ising variables $(q=2)$ to individuals based on opinions and inclinations. Frustrated closed loops of interactions will then lead to chaotic rescaling behavior, which implies large instabilities with respect to small changes in external parameters.  Perhaps this is artificial, as individuals can be in more than two possible states, and chaos can be avoided as seen in this paper.

\begin{acknowledgments}
Support by the Academy of Sciences of Turkey (T\"UBA) is gratefully
acknowledged.
\end{acknowledgments}

\end{document}